\documentclass{PoS}

\title{Magnetism in Cold-Dense QCD}

\ShortTitle{Magnetism in QCD}

\author{\speaker{Efrain J. Ferrer}\thanks{I want to thank the Organizing Committee of the 2011 Brazilian Workshop on Nuclear Physics for their kind invitation. The works that serve as the basis for this talk have been supported in part by DOE Nuclear Theory grant DE-SC0002179}\\
        Author affiliation\\Department of Physics, University of Texas at El Paso, 500 W. University Ave., El Paso, TX 79968, USA
        E-mail: \email{ejferrer@utep.edu}}

\abstract{In this talk I review some of the main findings on magnetism in color superconductivity. The physical characteristic of the different phases that are reached by increasing the applied magnetic field in a three-flavor color superconductor at asymptotically high densities are discussed. A mechanism to boost a seed magnetic field in high-dense quark matter is presented. Also, it is shown how a magnetic field can be generated in a color superconductor at moderate densities. Possible implications of these results for the astrophysics of magnetized compact objects are indicated.}

\FullConference{XXXIV edition of the Brazilian Workshop on Nuclear Physics,\\
		5-10 June 2011\\
		Foz de Iguaçu, Parana state, Brasil}

\begin{document}

\section{Introduction}
Color superconductivity (CS) was initially developed in the 70-80's \cite{CS} after the discovery of the phenomenon of asymptotic freedom in cold and dense QCD \cite{A-F}. Later on, CS regained a lot of interest when large pairing gaps \cite{CFL} and multiple phases with different color-charged condensates \cite{CFL}-\cite{Strong-2SC} were found in the context of effective models of low-energy QCD.

When matter is squeezed to densities several times higher than nuclear matter densities, free quarks are released thanks to the mechanism of asymptotic freedom. The ground state of the superdense quark system, a Fermi liquid of weakly interacting quarks, is unstable with respect to the formation of diquark condensates \cite{CS, CFL}, a nonperturbative phenomenon essentially equivalent to the Cooper instability of BCS superconductivity. In QCD, one gluon exchange between two quarks is attractive in the color-antitriplet channel. Thus, at densities much higher than the temperature, quarks condense into Cooper pairs, which are color antitriplets. These color condensates break the SU(3) color gauge symmetry of the ground state producing a color superconductor. Therefore, at zero temperature and sufficiently high densities, quark matter is in a CS state. At densities much higher than the masses of the $u$, $d$, and $s$ quarks, one can assume that the three quarks are massless. In this asymptotic region the favored state results to be the so-called Color-Flavor-Locking (CFL) state \cite{CFL}, characterized by a spin-zero diquark condensate antisymmetric in both color and flavor.

The stable CFL phase realized at high densities breaks down at intermediate densities due to the mismatch between the Fermi momenta of different quarks forming the Cooper pairs. This mismatch is produced by the strange quark mass $M_s$ and the constraints imposed by electric and color neutralities \cite{gCFL}. That is, although the validity of the CFL phase at asymptotically large densities is well settled, the next phase down in density is still a puzzle. The problem is that as a consequence of the mismatch in the Fermi momenta of the quarks forming the Cooper pairs the spectrum of some of the gluons exhibits chromomagnetic instabilities \cite{Inst-CFL}. This result posts a question about the system ground state at moderate densities. In the quest for the true ground state at this intermediate density region, several approaches have been considered. Among them we can mention: 1) the consideration of momentum-dependent condensates like the crystalline or LOFF phases \cite{CFL-Loff}; 2) the development of an effective theory of low-energy degrees of freedom that includes flavor rotations of the CFL condensate \cite{K-CFL}, and 3) the vortex formation of an inhomogeneous charged gluon condensate that induces magnetic flux tubes associated to the vortices \cite{Gluon-C}. However, none of the proposed solution has been proved to be the final answer and the question of the phase at intermediate densities still remains open. At even lower densities, where the strong coupling is strengthen, it has been found that the strange quark decouples leading to a stable two-flavor color superconducting phase \cite{Strong-2SC}.

The combination of high densities and relative low temperatures at which color superconducting Cooper pairs can form and resist the evaporation due to thermal effects could exist in the high dense cores of compact stars. This is so because the cores of neutron star remnants from supernovae explosion have densities several times larger than the saturation density of nuclear matter and temperatures several orders smaller than the superconducting gap. Thanks to the asymptotic freedom at those densities, the bulk of the system is governed by a relatively weak coupling. Hence, the physics of the core of compact stars is determined by the quark-gluon degrees of freedom, and the phase at the core could be one of the yet-to-be-determined color superconducting phases.

Compact stars, on the other hand, are typically strongly magnetized objects. Specifically, the so called
magnetars, can reach surface magnetic fields as large as $10^{14}-10^{15}$ G \cite{Magnetars}. Because the stellar medium has a very high electric conductivity, the magnetic flux should be conserved, thus, it is reasonable to expect larger magnetic field strengths in the regions of larger matter density, i.e. the star core. However, the interior magnetic fields of neutron stars are not directly accessible to observation, so their possible values can only be estimated with the help of heuristic methods. Estimates based on macroscopic and microscopic analysis, for nuclear \cite{virial}, and quark matter, considering both gravitationally bound and self-bound stars \cite{EoS-H}, have led to maximum fields within the range $10^{18}-10^{20}$ G, depending on the nature of the inner medium, that is, if it is formed by neutrons \cite{virial}, or quarks \cite{EoS-H}.

Contrary to what our na\"{\i}ve intuition might indicate, a magnetic field does not need to be of the order of the baryon chemical potential to produce a noticeable effect in a color superconductor. As shown in \cite{Phases, SpinoneCFL}, the color superconductor is characterized by various scales and different physics can occur at field strengths comparable to each of these scales. Specifically, as we will discuss in this talk, the superconducting CFL gap, the Meissner mass of the charged gluons and the baryon chemical potential define three order parameters that determine the values of the magnetic field needed to produce different effects in CS. That is, the presence of a sufficiently strong field in the star core can in principle modify the properties of the matter phase there and lead to observable signatures. Therefore, the investigation of the properties of very dense matter in the presence of strong magnetic fields is of interest not just from a fundamental point of view, but it is also closely connected to the physics of strongly magnetized neutron stars.

In this talk I review first the present knowledge of the magnetic effects in CS at asymptotically high densities where it is expected the realization of the CFL phase at zero applied field. In the second part, I show how a magnetic field can be generated in CS at moderate densities. Finally, I discuss some possible consequences of the existence of strongly magnetized cold-dense matter for astrophysics.

\section{The MCFL phase}

An important aspect of color superconductivity is to understand that it can be modified, without breaking the pairs, by the presence of a magnetic field. In a conventional superconductor, since Cooper pairs are electrically charged, the electromagnetic gauge invariance is spontaneously broken, thus producing a massive photon that can screen a weak magnetic field: the so called Meissner effect. In spin-zero color superconductivity, although the color condensate has non-zero electric charge, there is a linear combination of the photon and a the $8^{th}$ gluon field
\begin{equation}
{\tilde A}_\mu=\cos\theta_{CFL} A_\mu-\sin\theta_{CFL} G^8_\mu
\end{equation}
that remains massless \cite{rotated-EM}. The field $\tilde A_{\mu}$ plays the role of an in-medium or rotated electromagnetic field, as the color condensate is neutral with respect to the corresponding rotated charge. Thus, a magnetic field associated to $\tilde A_{\mu}$ can penetrate the CS without being subject to the Meissner effect. The orthogonal combination ${\tilde G}^8_\mu=\sin\theta_{CFL} A_\mu+\cos\theta_{CFL} G^8_\mu$ will be still massive. Actually, the penetrating field in the CFL superconductor is mostly formed by the original photon with only a small admixture of the 8th gluon since the mixing angle, $\cos^{-1}\theta_{CFL}=g/\sqrt{e^2/3+g^2}$, is sufficiently small.

The unbroken $\widetilde U(1)$ symmetry corresponding to the long-range rotated photon is generated by ${\tilde Q}=Q\times 1+1\times T_8/\sqrt{3}$, where $Q$ is the conventional electromagnetic charge of quarks and $T_8$ is the 8th Gell-Mann matrix. Using the matrix representations $Q=diag(-1/3,-1/3,2/3)$ for $(s,d,u)$ flavors and $T_8=diag(-1/\sqrt{3},-1/\sqrt{3},2/\sqrt{3})$ for $(b,g,r)$ colors, the ${\tilde Q}$ charges (in units of ${\tilde e}=e\cos\theta$) of different quarks are
\begin{equation}\label{table}
\begin{tabular}{ccccccccc}
\hline
\textrm{$s_b$}&
\textrm{$s_g$}&
\textrm{$s_r$}&
\textrm{$d_b$}&
\textrm{$d_g$}&
\textrm{$d_r$}&
\textrm{$u_b$}&
\textrm{$u_g$}&
\textrm{$u_r$}\\
0 & 0 & $-1$ & 0 & 0 & $-1$ & $+1$ & $+1$ & 0\\
\hline
\end{tabular}
\end{equation}

The first effect that an applied magnetic field can produce in a CS is that the gap structure gets modified due to the penetrating field \cite{MCFL}. To understand this, notice that, although the condensate is $\widetilde{Q}$-neutral, some of the quarks participating in the pairing are $\widetilde{Q}$-charged and hence can couple to the background $\widetilde{B}$-field, which in turn will affect the gap equations through the Green functions of these $\widetilde{Q}$-charged quarks. Due to the coupling of the charged quarks with the external field, the color-flavor space is augmented by the $\widetilde{Q}$-charge color-flavor operator, and consequently the CFL order parameter splits in new independent pieces giving rise to a new phase that we called Magnetic Color-Flavor-Locked (MCFL) phase \cite{MCFL}.

The symmetries in the MCFL phase are quite different from those in the CFL phase. To begin with, in the presence of an external magnetic field the flavor symmetries of QCD are reduced, as only the $d$ and $s$ quarks have equal electromagnetic charge. Then, the initial symmetry of the theory in the presence of a magnetic field is $SU(3) \times SU(2)_L \times SU(2)_R \times U(1)_B \times U^{(-)}(1)_A \times U(1)_{e.m.}$, where the $U^{(-)}(1)_A$ is connected with an anomaly free linear combination of the $u$, $d$ and $s$ axial currents \cite{miransky-shovkovy-02}. The antsatz used in \cite{MCFL} for the order parameter of the MCFL phase
\begin{equation}
\Phi^{\alpha\beta}_{ij}=\Delta\epsilon^{\alpha\beta 3}\epsilon_{ij3}+\Delta_B(\epsilon^{\alpha\beta 1}\epsilon_{ij1}+\epsilon^{\alpha\beta 2}\epsilon_{ij2})
\label{generalansatz}
\end{equation}
has a structure whose symmetry led to maximal unbroken symmetry in the color superconducting phase in the presence of a magnetic field. In (\ref{generalansatz}), $\Delta$ and $\Delta_B$ are the gap parameters, corresponding to Cooper pairs formed by neutral quarks and by neutral and charged quarks respectively. One can see that the MCFL order parameter implies the following symmetry breaking pattern: $SU(3) \times SU(2)_L \times SU(2)_R \times U(1)_B \times U^{(-)}(1)_A \times U(1)_{e.m.} \longrightarrow SU(2)_{C+L+R} \times \widetilde{U}(1)-{e.m}$, so it leaves a locked SU(2) symmetry group in addition to the long-range symmetry of the rotated electromagnetism. The symmetry breaking pattern in MCFL leads to a different low-energy physics than in the CFL phase \cite{Phases}.

In addition to changing the structure of the color superconductor, the magnetic field also affects the magnitude of the condensate, producing a very interesting and unique effect: the quark-quark condensation of pairs of quarks with nonzero rotated-charge is enhanced by the rotated magnetic field, so the magnitude of these condensates increases with the strength of the field \cite{MCFL}. This is the first example of a physical system where a magnetic field lends a hand to a superconductor.

An intuitive way to understand how a magnetic field can strengthen the condensation of pairs of quarks with equal and opposite rotated charges, while it weakens those formed by electron-electron pairs in conventional superconductivity is as follows. In a conventional superconductor, the electron-electron pair is formed by particles of equal charges and opposite spins, hence opposite magnetic moments. The field effect is to rotate one of the moments so to align both of them along the magnetic field direction, an effect that tends to break the condensate. In contrast, for the quark-quark pair, the magnetic moments of the quarks are parallel to each other; as the quarks of the pair have opposite spins and equal and opposite rotated charges. The field here tends to keep the magnetic moments pointing along the same direction; that is, it reinforces the original relative alignment of the moments thereby strengthening the pairing. Of course, this simple argument does not encompass all the complexity of the mechanism that ultimately led to a larger magnitude for the condensate of pairs formed by rotated-charged quarks \cite{MCFL}.

As it is apparent from the above intuitive argument, the situation in CS has some resemblance to the magnetic catalysis of chiral symmetry breaking \cite{MC, MC-AMM}, in the sense that the magnetic field strengthens the pair formation of pairs of particles with opposite spins and opposite charges. Despite this similarity, it can be easily seen that the way the field influences the pairing mechanism in the two cases is quite different. The particles participating in the chiral condensate are near the surface of the Dirac sea. The effect of a magnetic field there is to effectively reduce the spatial dimension for the particles lying at the lowest Landau level (LLL), which in turn strengthen their effective coupling, catalyzing the chiral condensate. Color superconductivity, on the other hand, involves quarks near the Fermi surface, with a pairing dynamics that is already (1+1)-dimensional. Therefore, the $\widetilde{B}$-field does not yield further dimensional reduction of the pairing dynamics near the Fermi surface and hence the LLL does not play any special role in the color superconductor. What the field does in the color superconductor is to increase the density of states of the $\widetilde{Q}$-charged quarks, and it is through this mechanism that the pairing of the charged particles is reinforced by the penetrating magnetic field, producing larger gaps.

Natural questions to be asked at this point are: How can be physically distinguished the MCFL phase from the CFL one? and, is there a threshold field to produce the MCFL phase?

To answer these questions we should take into account that the main distinctive characteristic of each phase is given by the number of Goldstone bosons that corresponds to their different global symmetries.
In the CFL phase the symmetry breaking is given
by $SU(3)_C \times SU(3)_L \times SU(3)_R \times U(1)_B
\times U(1)_{\rm e.m.}\rightarrow SU(3)_{C+L+R}\times {\widetilde
U(1)}_{\rm e.m.}$. This symmetry reduction leaves nine Goldstone bosons: a singlet
associated to the breaking of the baryonic symmetry $U(1)_B$, and
an octet associated to the axial $SU(3)_A$ group. While as a consequence of the symmetry breaking pattern of the MCFL phase we have that only five
Goldstone bosons remain. Three of them correspond to the breaking
of $SU(2)_A$, one to the breaking of $U(1)^{(-)}_A$, and one to
the breaking of $U(1)_B$. Thus, an applied magnetic field reduces
the number of Goldstone bosons in the superconducting phase, from
nine to five \cite{Phases}. Moreover, the MCFL phase is not just characterized by a smaller number of
Goldstone fields, but by the fact that all its bosons are
neutral with respect to the rotated electric charge. Hence, no
charged low-energy excitation can be produced in the MCFL phase. Moreover, in the MCFL phase, as well as in the CFL one, the fermion excitations are gapped, and the gluon fields acquire masses thanks to the Meissner-Anderson-Higgs mechanism. Hence, the MCFL phase behaves as an insulator, as it has no low-energy charged excitations at zero temperature.

The neutral low-energy particle spectrum of the MCFL phase can be relevant for the low energy physics of color
superconducting star cores and hence for its transport
properties. In particular, the cooling of a compact star is
determined by the particles with the lowest energy; so a star with
a core of quark matter and sufficiently large magnetic field can
have a distinctive cooling process.

Regarding the second question, we have that in principle once a magnetic field is
present the symmetry is theoretically that of the MCFL, as
discussed above. However, in practice for $\widetilde{B}<
\widetilde{B}_{MCFL}\sim \Delta^{2}_{CFL}$ the MCFL phase is
almost indistinguishable from the CFL. Only at fields comparable
to $\Delta^{2}_{CFL}$ the main features of MCFL emerge through the
low-energy behavior of the system \cite{Phases}. At the threshold field
$\widetilde{B}_{MCFL}$, only five neutral Goldstone bosons remain
out of the original nine characterizing the low-energy behavior of
the CFL phase, because the charged Goldstones acquire field dependent masses and can decay in lighter modes.
For a meson to be stable in this system, its mass
should be less than twice the gap, otherwise it will decay into a
particle-antiparticle pair. That means that, as proved in Ref. \cite{Phases}, once the applied field produces a mass for the charged Goldstones of the order of the CFL gap it is reached the threshold field
for the effective $CFL \rightarrow MCFL$ symmetry transmutation. Therefore, coming from low to higher fields, the first magnetic phase that will effectively show up in the magnetized system will be
the MCFL, even though at fields near the threshold field the
splitting of the gaps may still be negligible \cite{MCFLoscillation}.

The neutral five Goldstone bosons are the ones determining the new low-energy behavior of the genuinely realized
MCFL phase. Hence, in a CS with three-flavor quarks at
very high densities an increasing magnetic field produces a phase
transmutation from CFL to MCFL at a threshold field proportional to the square of the CFL gap ($\widetilde{B} \simeq
\Delta^{2}_{CFL}$). During the phase transmutation no symmetry
breaking occurs, since in principle once a magnetic field is present
the symmetry is theoretically that of the MCFL phase. The existence of this phase transmutation is manifested in the behavior of the gaps versus the magnetic field. At field strength smaller than the threshold field we find that $\Delta \approx \Delta_B \approx \Delta_{CFL}$, while for larger fields the gaps exhibit oscillations with respect to ${\tilde e}{\tilde B}/\mu^2$ \cite{MCFLoscillation}, owed to the de Haas-van Alphen effect \cite{HvA}.

Recently, it was also found that in the MCFL phase a new condensate associated with the magnetic moment of the Cooper pairs is realized \cite{SpinoneCFL}. It should be taken into account that, as we discussed previously, in this phase the Cooper pairs formed by charged quarks have nonzero magnetic moment, since the quarks in the pair no only have opposite charge but also opposite spin. Hence, as found in Ref. \cite{SpinoneCFL}, the magnetic moment of this kind of pairs leads in principle to a nonzero net magnetic moment for the system, which in turn would be reflected in the existence of an extra condensate $\Delta_M$ in addition to the $\Delta$ and $\Delta_{B}$ gaps.

Symmetry arguments can give additional insight on the existence of this extra gap parameter $\Delta_M$ in the MCFL phase. The presence of a magnetic field breaks the spatial rotational symmetry $O(3)$ to the subgroup of rotations $O(2)$ about the axis parallel to the field. This symmetry breaking opens new attractive pairing channels through the new Fierz identities that were not available in the CFL phase. One of these channels has Dirac structure $\Delta_M\sim C\gamma_5\gamma^1\gamma^2$. Then, to ensure the total antisymmetry required by the Pauli principle, and given the symmetric nature of the new gap under transposition in Dirac indexes, together with its color antisymmetry, which guarantees the strongest attractive interaction in color, it was proposed a new gap symmetric in flavor to be added to (\ref{generalansatz})
\begin{equation}
\overline{\Phi}^{\alpha\beta}_{ij}=\Delta_M([\epsilon^{\alpha\beta 1}(\delta_{i2}\delta_{j3}+\delta_{i3}\delta_{j2})+\epsilon^{\alpha\beta 2}(\delta_{i1}\delta_{j3}+\delta_{i3}\delta_{j1})],
\label{generalansatz-2}
\end{equation}

A condensate with this structure corresponds to a magnetic moment condensate. Because it does not break any symmetry that has not already been broken by the gaps $\Delta$ and $\Delta_{B}$, a magnetic moment condensate is not symmetry-protected and in principle should exit. As we found in Ref. \cite{SpinoneCFL}, this new condensate is spin-one and since it is a direct consequence of the external magnetic field its magnitude becomes comparable to the energy gaps only at strong field values. Although this new condensate vanishes at zero magnetic field, we can not ignore it even at very small magnetic fields, because the gap-equation solution $\Delta\neq 0$, $\Delta_B\neq 0$, $\Delta_M=0$ is not allowed. Thus, we have to take into account $\Delta_M$ at equal footing with the other gaps once the magnetic field is different from zero.

The magnetic-moment condensate of the MCFL phase shares a few similarities with the dynamical generation of an anomalous magnetic moment recently found in massless QED \cite{MC-AMM}. Akin to the Cooper pairs of oppositely charged quarks in the MCFL phase, the fermion and antifermion that pair in massless QED also have opposite charges and spins and hence carries a net magnetic moment. A dynamical magnetic moment term in the QED Lagrangian does not break any symmetry that has not already been broken by the chiral condensate. Therefore, once the chiral condensate is formed due to the magnetic catalysis of chiral symmetry breaking \cite{MC, MC-AMM}, the simultaneous formation of a dynamical mass and a dynamical magnetic moment is unavoidable \cite{MC-AMM}.

If we represent the gaps obtained from the solutions of the three gap equations, $\partial\Omega/\partial \Delta = 0$, $\partial\Omega/\partial \Delta_B = 0$, and $\partial\Omega/\partial \Delta_M = 0$, versus $\widetilde{e}\widetilde{B}/\mu^2$, with $\Omega$ being the free energy of the MCFL phase, we obtain the graphs represented in Fig. 1 \cite{SpinoneCFL}.

\begin{figure}
\includegraphics[width=.9\textwidth]{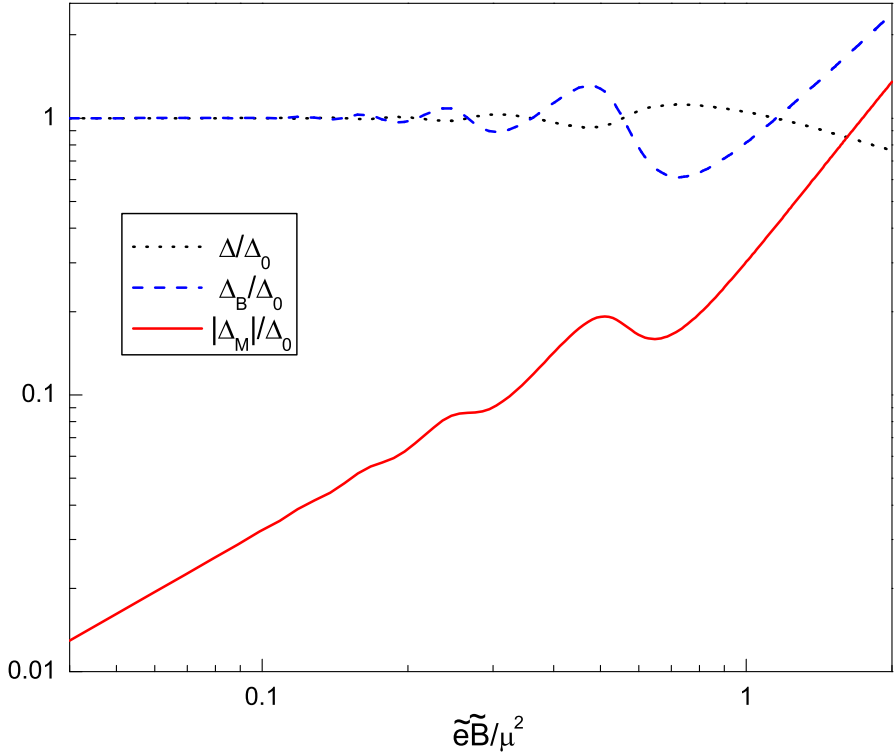}
\caption{The three gaps of the MCFL phase as a function of ${\tilde e}{\tilde B}/\mu^2$ for $\mu=500MeV$.}
\label{fig1}
\end{figure}

It is apparent from the graphical representation of $\Delta_M$ in Fig. 1, that its value remains relatively small up to magnetic-field values of the order of $\mu^2$; and contrary to the other $\Delta's$ gaps, without signals of de Hass-van Alphen oscillations. This is due to the fact that only Cooper pairs formed by particles in the LLL contribute to the magnetic moment condensate. Higher LL's ($l>1$) cannot contribute since they accommodate particles having the same charge and opposite spins, thus contributing to the formation of Cooper pairs with opposite magnetic moments. At low fields, the filling of the LLL is scarce, while for fields of order $\widetilde{e}\widetilde{B}\geq \mu^2$, all the particles are restrained to the LLL, hence the variation of $\Delta_M$ from lower values at weak field, to higher values at sufficiently strong fields. The absence of de Hass-van Alphen oscillations is due to the fact that there is no change in LL's contributing to $\Delta_M$.

The difference found between the gaps $\Delta$ and $\Delta_B$ with $\Delta_M$ is also in agreement with the fact that the induced expectation value of the magnetic moment is a quantity purely generated by the magnetic field (i.e. $\Delta_M=0$ at $B=0$), while the relevant scale for the generation of $\Delta$ and $\Delta_B$ is the energy at the Fermi surface, i.e. the chemical potential $\mu$. Therefore, once the magnitude of the magnetic field is as large as $\mu$, the induced average magnetic moment becomes as large as the gap containing charged quarks $\Delta_B$. Otherwise, the magnetic moment is negligibly small as compared with the other gaps. Another important consequence of the generation of the average magnetic moment is that its presence strengthens the gap $\Delta_B$ in the sufficiently strong-magnetic-field region, as can be checked by comparing our results in Fig. 1 with those of Ref. \cite{MCFLoscillation}.

\section{The Paramagnetic-CFL phase}

Now, we analyze how the gluons are affected by an applied magnetic field in a CS state and how at sufficiently strong magnetic fields a new phase, that we called the Paramagnetic-CFL (PCFL) phase \cite{Vortex}, is created. It is known that gluons are neutral with respect to the conventional electromagnetism, but some of them acquire rotated electric charges in a color superconductor. In the CFL phase the $\widetilde{Q}$-charge of the gluons in units of $\widetilde{e}$ are
\begin{equation}\label{table-2}
\begin{tabular}{cccccccc}
\hline
\textrm{$G_{\mu}^{1}$}&
\textrm{$G_{\mu}^{2}$}&
\textrm{$G_{\mu}^{3}$}&
\textrm{$G_{\mu}^{+}$}&
\textrm{$G_{\mu}^{-}$}&
\textrm{$I_{\mu}^{+}$}&
\textrm{$I_{\mu}^{-}$}&
\textrm{$\widetilde{G}_{\mu}^{8}$}\\
0 & 0 & $0$ & 1 & -1 & $1$ & $-1$ & 0\\
\hline
\end{tabular}
\end{equation}
The $\widetilde{Q}$-charged fields in (\ref{table-2}) correspond to the
combinations $G_{\mu}^{\pm}\equiv\frac{1}{\sqrt{2}}[G_{\mu}^{4}\mp
iG_{\mu}^{5}]$ and
$I_{\mu}^{\pm}\equiv\frac{1}{\sqrt{2}}[G_{\mu}^{6}\mp iG_{\mu}^{7}]$. The $\widetilde{G}_{\mu}^{8}$ field is the short-range rotated field defined in the previous section.

To investigate the effect of the applied rotated magnetic field
$\widetilde{H}$ on the charged gluons, we should start from the effective action of the charged fields $G_{\mu}^{\pm}$ (the
contribution of the other charges gluons $I_{\mu}^{\pm}$ is similar)
\begin{eqnarray}
\label{Eff-Act-3} \Gamma_{eff}& = & \int dx
\{-\frac{1}{4}(\widetilde{f}_{\mu
\nu})^{2}+G_{\mu}^{-}[(\widetilde{\Pi}_{\mu}\widetilde{\Pi}_{\mu})\delta_{\mu
\nu}-2i\widetilde{e}\widetilde{f}_{\mu \nu}\nonumber
 \\
& - & (m_{D}^{2} \delta_{\mu 0} \delta_{\nu 0}+ m_{M}^{2}
\delta_{\mu i} \delta_{\nu i})-(1-\frac{1}{\varsigma}
\widetilde{\Pi}_{\mu}\widetilde{\Pi}_{\nu})]G_{\nu}^{+}\}
\end{eqnarray}
Here, $\varsigma$ is the gauge fixing parameter, $\widetilde{\Pi}_{\mu}=\partial_{\mu}
-i\widetilde{e}\widetilde{A}_{\mu}$ is the covariant derivative in the presence of the external rotated field, $m_{D}$ and $m_{M}$ are the $G_{\mu}^{\pm}$-field Debye and Meissner masses respectively, and the field strength tensor for the rotated electromagnetic field if denoted by $\widetilde{f}_{\mu
\nu}=\partial_{\mu}\widetilde{A}_{\nu}-\partial_{\nu}\widetilde{A}_{\mu}$.
 The corresponding Debye and Meissner masses in (\ref{Eff-Act-3})
are given by \cite{Wang}
\begin{equation}
m_{D}^{2} = m_{g}^{2} \frac{21-8 \ln 2}{18},\qquad m_{M}^{2} =
m_{g}^{2} \frac{21-8 \ln 2}{54},
\end{equation}
with $m_{g}^{2}=g^2(\mu^{2}/2\pi^{2})$. We
are neglecting the correction produced by the applied field to the
gluon Meissner masses since it will be a second order effect. The effective action (\ref{Eff-Act-3}) is characteristic of a
spin-1 charged field in a magnetic field (for details see for
instance \cite{emilio}).

Assuming an applied magnetic field
along the third spatial direction
($\widetilde{f}^{ext}_{12}=\widetilde{H}$), we find after diagonalizing the
mass matrix of the field components ($G^{+}_{1}, G^{+}_{2}$) in (\ref{Eff-Act-3})
\begin{equation}
%\begin{eqnarray}
%\label{gapMCFL}
\left(
\begin{array}{cc}
m_{M}^{2}& i\widetilde{e}\widetilde{H} \\
- i\widetilde{e}\widetilde{H}& m_{M}^{2}
 \label{mass-matrx}
\end{array} \right) \rightarrow
\left(
\begin{array}{cc}
m_{M}^{2}+\widetilde{e}\widetilde{H}& 0 \\
0& m_{M}^{2}-\widetilde{e}\widetilde{H}
 \label{mass-matrx}
\end{array} \right),
%\end{eqnarray}
\end{equation}
with corresponding eigenvectors ($G^{+}_{1}, G^{+}_{2}$)
$\rightarrow$ ($G,iG$). We see that the lowest mass mode in (\ref{mass-matrx}) has a sort of
"Higgs mass" above the critical field
$\widetilde{e}\widetilde{H}_{C}= m_{M}^2$, indicating the setup of an instability for the
$G$-field. This phenomenon is the well known "zero-mode
problem" found in the presence of a magnetic field for Yang-Mills
fields \cite{zero-mode}, for the $W^{\pm}_{\mu}$ bosons in the
electroweak theory \cite{Skalozub, Olesen}, and even for
higher-spin fields in the context of string theories
\cite{porrati} and it is due to the presence of the gluon anomalous magnetic moment term $2i\widetilde{e}\widetilde{f}_{\mu
\nu}G_{\mu}^{-}G_{\nu}^{+} $ in (\ref{Eff-Act-3}).
Thus, to remove
the instability it is needed the restructuring of the ground
state through the condensate of the field bearing the tachyonic mode (i.e. the $G$-field).

We should stress that the gluon condensate discussed in this talk is not the
only charged spin-one condensate generated in a theory with a large fermion density.
As known \cite{Linde}, a spin-one condensate of W-bosons can be originated at sufficiently high
fermion density in the context of the electroweak theory at zero magnetic field. However,
the physical implications of the gluon condensate induced by the magnetic field in the
CS are fundamentally different from those associated to the homogeneous W-boson
condensate of the dense electroweak theory \cite{Linde}. One of the main physical differences is that the homogeneous W condensate, being electrically charged, so to compensate the excess of charge due to the finite density of electrons \cite{Linde}, breaks the electromagnetic $U(1)$ group producing a conventional superconducting state \cite{W-condensate}; while the inhomogeneous gluon condensate in CS is formed with gluons of both charges, so keeping the condensate state neutral.

To find the $G$-field condensate and the induced magnetic field
$\widetilde{\textbf{B}}=\nabla\times\widetilde{\textbf{A}}$, with
$\widetilde{\textbf{A}}$ being the total rotated electromagnetic
potential in the condensed phase in the presence of the external
field $\widetilde{H}$, we should start from the Gibbs free energy
density $\mathcal{G}=\mathcal{F}-\widetilde{H}\widetilde{B}$, since
it depends on both $\widetilde{B}$ and $\widetilde{H}$
($\mathcal{F}$ is the system free energy density). Since
specializing $\widetilde{H}$ in the third direction the instability
develops in the $(x,y)$-plane, we make the ansatz for the condensed field $\overline{G}=\overline{G}(x,y)$.
Starting from (\ref{Eff-Act-3}) in the Feynman gauge $\varsigma=1$,
which in terms of the condensed field $\overline{G}$ implies
$(\widetilde{\Pi}_{1}+i\widetilde{\Pi}_{2})\overline{G}=0$, we have that the
Gibbs free energy in the condensed phase is
\begin{equation}
\label{Gibbs} \mathcal{G}_{c} =\mathcal{F}_{n0} +\widetilde{\Pi}^{2}
\overline{G}^{2}-2(\widetilde{e}\widetilde{B}-m_{M}^{2})\overline{G}^{2}+2g^{2}\overline{G}^{4}+\frac{1}{2}\widetilde{B}^{2}-\widetilde{H}\widetilde{B}.
\end{equation}
where $\mathcal{F}_{n0}$ is the system free energy in the normal
phase ($\overline{G}=0$) at zero magnetic field.

The minimum equations for the fields $\overline{G}$ and
$\widetilde{B}$ are respectively obtained from (\ref{Gibbs}) as
\begin{equation}
\label{EqG} \widetilde{\Pi}^{2}
\overline{G}+2(m_{M}^{2}-\widetilde{e}\widetilde{B})\overline{G}+8g^{2}\overline{G}^{2}\overline{G}=0,
\end{equation}

\begin{equation}
\label{EqB} 2\widetilde{e} \overline{G}^{2}-\widetilde{B}+\widetilde{H}=0
\end{equation}
Identifying $\overline{G}$ with the complex order parameter, Eqs.
(\ref{EqG})-(\ref{EqB}) become analogous to the Ginzburg-Landau
equations for a conventional superconductor except by the negative sign in front of the $\widetilde{B}$ field in Eq. (\ref{EqG}) and the positive sign in the first term of the LHS of Eq. (\ref{EqB}) \cite{Vortex}. The fact that those signs turn the opposite of those appearing in conventional superconductivity is due to the different nature of the condensates in both cases. While in conventional superconductivity the Cooper pair is formed by spin-1/2 particles, here we have a condensate formed by spin-1 charged particles interacting through their anomalous magnetic moment with the magnetic field (i.e. the term $2i\widetilde{e}\widetilde{f}_{\mu
\nu}G_{\mu}^{-}G_{\nu}^{+} $ in (\ref{Eff-Act-3})).

Notice that because of the different sign in
the first term of (\ref{EqB}), contrary to what occurs in
conventional superconductivity, the resultant field $\widetilde{B}$
is stronger than the applied field $\widetilde{H}$. Thus, when a
gluon condensate develops, the magnetic field will be antiscreened
and the color superconductor will behave as a paramagnet. The
antiscreening of a magnetic field has been also found in the context
of the electroweak theory for magnetic fields $H \geq
M_{W}^{2}/e\sim 10^{24} G$ \cite{Olesen}. Just as in the electroweak
case, the antiscreening in the color superconductor is a direct
consequence of the asymptotic freedom of the underlying theory
\cite{Olesen}.

We conclude that the magnetic field in the new
phase is boosted to a higher
value, which depends on the modulus of the $\overline{G}$-condensate. That is why the phase attained at $\widetilde{H}\geq \widetilde{H}_c$ is called paramagnetic CFL (PCFL) \cite{Phases, Vortex}. It should be pointed out that at the scale of baryon densities
typical of neutron-star cores ($\mu \simeq 400 MeV$, $g(\mu)\simeq
3$) the charged gluons magnetic mass in the CFL phase is $m_{M}^{2}
\simeq 16\times 10^{-3} GeV^{2}$. This implies a critical magnetic
field of order $\widetilde{H}_{c}\simeq 0.7\times 10^{17} G$. Although it is a significant high value, it is in the expected range
for the neutron star interiors with cold-dense quark matter \cite{EoS-H}. Let us stress that in our analysis
we considered asymptotic densities where quark masses can be
neglected. Al lower densities where the Meissner masses of the charged gluons become smaller, the field values needed to develop the magnetic instability will be smaller.

To find the structure of the gluon condensate we should solve the non-linear differential equation (\ref{EqG}). However, to get an analytic solution we can consider the approximation where  $\widetilde{H}\approx \widetilde{H}_c=m_M^2$ and consequently $\mid \overline{G}\mid \approx 0$. In this approximation, Eq. (\ref{EqG}) can be linearized as
\begin{equation}
\label{EqVortex} [\partial_{j}^{2}-\frac{4\pi
i}{\widetilde{\Phi}_{0}}\widetilde{B}x\partial_{y}-4\pi^{2}\frac{\widetilde{B}^{2}}{\widetilde{\Phi}_{0}^{2}}x^{2}-\frac{1}{\xi^{2}}]\overline{G}=0,
\qquad j=x,y
\end{equation}
where we fixed the gauge condition
$\widetilde{A}_{2}=\widetilde{B}x_{1}$, and introduced the notations
$\widetilde{\Phi}_{0}=2\pi/\widetilde{e}$, and
$\xi^{2}=\frac{1}{2}(\widetilde{e}\widetilde{B}-m_{M}^{2})^{-1}$.

Eq. (\ref{EqVortex}) is formally similar to the Abrikosov's equation in type-II conventional superconductivity \cite{Abrikosov}, with $\xi$ playing the role of the coherence length and $\widetilde{\Phi}_{0}$ of the flux quantum per vortex cell. Then, following the Abrikosov's approach, a solution of Eq. (\ref{EqVortex}) can be found as
\begin{equation}
\label{Vortex-solution} \overline{G}(x,y)=\frac{1}{\sqrt{2}\widetilde{e}\xi}e^{-\frac{x^2}{e\xi^2}}\vartheta_3(u/\tau),
\end{equation}
with $\vartheta_3(u/\tau)$ being the elliptic theta function with arguments
\begin{equation}
\label{arguments} u=-i\pi b(\frac{x}{\xi^2}+\frac{y}{b^2}), \qquad \tau=-i\pi\frac{b^2}{\xi^2}
\end{equation}
In (\ref{arguments}) the parameter $b$ is the periodic length in the y-direction ($b=\Delta y$). The double periodicity of the elliptic theta function also implies that there is a periodicity in the x-direction given by $\Delta x=\widetilde{\Phi}_{0}/b\widetilde{H}_{c}$. Therefore, the magnetic flux through each
periodicity cell ($\Delta x \Delta y$) in the vortex lattice is quantized $\label{Flux}
\widetilde{H}_c \Delta x \Delta y=\widetilde{\Phi}_{0}$, with
$\widetilde{\Phi}_{0}$ being the flux quantum per unit vortex cell.
In this semi-qualitative analysis we considered the Abrikosov's
ansatz of a rectangular lattice, but the lattice configuration should be carefully
determined from a minimal energy analysis. For the rectangular
lattice, we see that the area of the unit cell is $A=\Delta x \Delta
y=\widetilde{\Phi}_{0} /\widetilde{H}_c$, so decreasing with
$\widetilde{H}$.

In conclusion, we have that to remove the instability
a magnetic field specialized along the $z$-direction turns
inhomogeneous in the $(x,y)$-plane since it depends on the
condensate $\overline{G}$, which has periodicity on that plane, while it can be
homogeneous in the $z$-direction, therefore it forms a fluxoid along
the $z$-direction that creates a nontrivial topology on the
perpendicular plane. From (\ref{B-Eq}) we see that the magnetic
field can go from a minimum value $\widetilde{H}$ to a maximum at
the core of the fluxoid that depends on the amplitude of the gluon
condensate determined by the mismatch between the applied field and
the gluon Meissner mass.

Summarizing, at low $\widetilde{H}$ field, the CFL phase behaves as an
insulator, and the $\widetilde{H}$ field just penetrates through it without any change of strength.
At sufficiently high $\widetilde{H}$, the condensation of $G^{\pm}$
is triggered inducing the formation of a lattice of magnetic flux
tubes that breaks the translational and remaining rotational
symmetries. It should be noticed that contrary to the situation in conventional type-II
superconductivity, where the applied field only penetrates through the
flux tubes and with a smaller strength, the vortex state in the color superconductor has the peculiarity that outside the flux tube the
applied field $\widetilde{H}$ totally penetrates the sample, while
inside the tubes the magnetic field becomes larger than
$\widetilde{H}$. This effect provides an internal mechanism to increase the magnetic field of a compact star with a CS core.

\section{Chromomagnetic instability and induced magnetism in CS}

As discussed above, chromomagnetic instabilities can be present in CS even in
the absence of an external magnetic field. These instabilities may
appear at moderate densities after imposing electrical and color neutralities, as well as
$\beta$ equilibrium conditions, and at densities where the $s$
quark mass $M_{s}$ becomes a relevant parameter. As found first in
$g2SC$ \cite{Igor}, and then also in $gCFL$ \cite{Inst-CFL}, some
charged gluons typically become tachyonic at the onset of the
gapless phase. Here, I will discuss how the chromomagnetic instabilities in the 2SC system in the absence of an applied magnetic field can be removed by the spontaneous generation of a condensate of inhomogeneous gluons that simultaneously induces a rotated magnetic field. It is expected that a similar mechanism can be also found for the unstable CFL phase although it is a pending task.

In the 2SC phase the gluons' charges in units of $\widetilde{e} =
e \cos{\theta_{2SC}}$, with $\cos^{-1}{\theta_{2SC}}=\sqrt{1+\frac{1}{3}(\frac{e}{g})^2}$, are

\begin{equation}\label{table-3}
\begin{tabular}{cccccc}
\hline
\textrm{$G_{\mu}^{1}$}&
\textrm{$G_{\mu}^{2}$}&
\textrm{$G_{\mu}^{3}$}&
\textrm{$K_{\mu}$}&
\textrm{$K_{\mu}^{\dag}$}&
\textrm{$\widetilde{G}_{\mu}^{8}$}\\
0 & 0 & $0$ & 1/2 & -1/2 & 0\\
\hline
\end{tabular}
\end{equation}
where we used for the charged fields the doublet representation
\begin{eqnarray}  \label{charged-fields-2SC}
K_{\mu} \equiv \frac{1}{\sqrt{2}}\left(
\begin{array}{cc}
G_{\mu}^{(4)}-iG_{\mu}^(5)\\
G_{\mu}^{(6)}-iG_{\mu}^(7)
\end{array}
\right) \
\end{eqnarray}
The charged gluon fields $K_{\mu}^{\pm}$ can interact, through the
long-range field $\widetilde{A}_{\mu}$, with an applied external
magnetic field.

In the gapped $2SC$ phase the solution of the neutrality conditions
$\partial \Omega_{0}/\partial \mu_{i}=0$, with $\mu_{i}=\mu_{e},
\mu_{8}, \mu_{3}$, and gap equation $\partial \Omega_{0}/\partial
\Delta=0$, for the system free energy $\Omega_{0}$ in the
mean-field approximation, led to $\mu_{3}=0$, and nonzero values of
$\mu_{e}$, and $\mu_{8}$, satisfying $\mu_{8}\ll \mu_{e}<\mu$ for a
wide range of parameters \cite{Huang}. Here $\mu$ is the quark
chemical potential, $\mu_{e}$ the electric chemical potential, and
the "chemical potentials" $\mu_{8}$ and $\mu_{3}$ are strictly
speaking condensates of the time components of gauge fields,
$\mu_{8}= (\sqrt{3}g/2)\langle G_{0}^{(8)}\rangle$ and $\mu_{3}=
(g/2)\langle G_{0}^{(3)}\rangle$. The nonzero values of the chemical
potentials produce a mismatch between the Fermi spheres of the quark
Cooper pairs, $\delta \mu=\mu_{e}/2$.

The gapped 2SC turned out to be unstable once the gluon fields were
taken into consideration. As shown in Ref. \cite{Igor} by calculating the corresponding polarization operators in the CS phase under the neutrality and $\beta$ equilibrium conditions, the gluons
$G_{\mu}^{(1,2,3)}$ are massless, the in-medium $8^{th}$-gluon
$\widetilde{G}_{\mu}^8$
has positive Meissner square mass, and the $K$-gluon doublet
has Meissner square mass that becomes imaginary for $\Delta > \delta
\mu > \Delta/\sqrt{2}$, signalizing the onset of an unstable ground
state. The mass of the in-medium (rotated) electromagnetic field
$\widetilde{A}_{\mu}$
is zero, which is consistent with the remaining unbroken
$\widetilde{U}(1)_{em}$ group.

Similarly to the case of an external field analyzed in the last section, to study the condensation phenomenon triggered by the
tachyonic modes of the charged gluons, we consider
the gauge sector of the mean-field effective action
that depends on the charged gluon fields
and rotated electromagnetic field. For a static solution, one only
needs the leading contribution of the polarization operators in the
infrared limit ($p_{0}=0, |\overrightarrow{p}|\rightarrow 0$). Under
these conditions, the effective action of the charged gluons in interaction with a rotated magnetic field can be
written as
\begin{eqnarray}
\label{Eff-Act-2} \Gamma_{eff}^{g}& = & \int d^{4}x
\{-\frac{1}{4}(\widetilde{f}_{\mu
\nu})^{2}-\frac{1}{2}|\widetilde{\Pi}_{\mu}K_{\nu}-\widetilde{\Pi}_{\nu}K_{\mu}|^{2}
\nonumber
 \\
& - & [ m_{M}^{2} \delta_{\mu i} \delta_{\nu
i}-(\mu_{8}-\mu_{3})^{2} \delta_{\mu \nu}+
i\widetilde{q}\widetilde{f}_{\mu \nu}] K_{\mu}K_{\nu}^{\dag}\qquad
\nonumber
 \\
 & + &
\frac{g^2}{2}[(K_{\mu})^{2}(K^{\dag}_{\nu})^{2}-(K_{\mu}K^{\dag}_{\mu})^{2}]
+\frac{1}{\lambda}K^{\dag}_{\mu}\widetilde{\Pi}_{\mu}\widetilde{\Pi}_{\nu}K_{\nu}\}\qquad
\end{eqnarray}
where
$m_{M}^{2}=\frac{2\alpha_{s}\overline{\mu}^{2}}{3\pi}[1-\frac{2\delta\mu^{2}}{\Delta^{2}}]$,
is the Meissner mass with $\overline{\mu}=\mu -\frac{\mu_{e}}{6} +
\frac{\mu_{8}}{3}$ and $\alpha_{s}\equiv\frac{g^{2}}{4\pi}$
\cite{Igor},
$\widetilde{\Pi}_{\mu}=\partial_{\mu}
-i\widetilde{q}\widetilde{A}_{\mu}$ and $\widetilde{f}_{\mu
\nu}=\partial_{\mu}\widetilde{A}_{\nu}-\partial_{\nu}\widetilde{A}_{\mu}$.
In (\ref{Eff-Act-2}) the Debye mass $m_{D}$ was not included since
it will be no substantial for our derivations. The chemical
potential $\mu_{3}$, although is zero in the gapped phase, should be
in principle taken into account in the analysis of the new phase,
since a K-condensate breaks the remaining $SU(2)_c$ symmetry.

As usual in theories with zero-component gauge-field condensates \cite{Linde},
$\mu_{8}$ gives rise to a tachyonic mass contribution. Thus, coming
from $\delta \mu < \Delta/\sqrt{2}$ (where $m^{2}_{M}>0$), when
$m^{2}_{M}-\mu^{2}_{8} < 0$ a tachyonic mode develops. Borrowing
from the experience gained in the case with external magnetic field \cite{Vortex} (see previous section),
we expect that this instability should also be removed through the
spontaneous generation of an inhomogeneous gluon condensate $\langle
K_{i} \rangle$ capable to induce a rotated magnetic field, thanks to
the anomalous magnetic moment associated to these spin-1 charged particles.
Having this in mind, it was proposed the following ansatz \cite{Gluon-C}
\begin{eqnarray}  \label{condensate}
\langle K_{\mu} \rangle \equiv \frac{1}{\sqrt{2}}\left(
\begin{array}{cc}
\overline{G}_{\mu}\\
0
\end{array}
\right) \ , \quad \overline{G}_{\mu} \equiv
\overline{G}(x,y)(0,1,-i,0),
\end{eqnarray}
where it was taken advantage of the $SU(2)_{c}$ symmetry to write the
$\langle K_{i} \rangle$-doublet with only one nonzero component.
Since in this ansatz the inhomogeneity of the gluon condensate is
taken in the $(x,y)$-plane, it follows that the corresponding
induced magnetic field will be aligned in the perpendicular
direction, i.e. along the z-axes, $\langle\widetilde{f}_{12}
\rangle=\widetilde{B}$. The part of the free-energy density that
depends on the gauge-field condensates,
$\mathcal{F}_{g}=\mathcal{F}-\mathcal{F}_{n0}$, with
$\mathcal{F}_{n0}=-\Gamma_{0}=\Omega_{0}$ denoting the system
free-energy density in the absence of the gauge-field condensate
($\overline{G}=0, \widetilde{B}=0$), is found, after fixing the
gauge parameter to $\lambda=1$ and using the ansatz
(\ref{condensate}) in (\ref{Eff-Act-2}), to be
\begin{eqnarray}
\label{free-energy} \mathcal{F}_{g} =
\frac{\widetilde{B}^{2}}{2}-2\overline{G}^{\ast}\widetilde{\Pi}^{2}
\overline{G}+2g^{2}|\overline{G}|^{4}\qquad\qquad \nonumber
\\
 -2[2\widetilde{q}\widetilde{B}+(\mu_{8}-\mu_{3})^2-m_{M}^{2}]|\overline{G}|^{2}\qquad
\end{eqnarray}
From the neutrality condition for the $3^{rd}$-color charge it is
found that $\mu_{3}=\mu_{8}$. The fact that $\mu_{3}$ gets a finite
value just after the critical point $m^{2}_{M}-\mu^{2}_{8} = 0$ is
an indication of a first-order phase transition, but since $\mu_{8}$
is parametrically suppressed in the gapped phase by the quark
chemical potential $\mu_{8}\sim \Delta^{2}/\mu$ \cite{Igor}, it will
be a weakly first-order phase transition. As follows, we will
consider that $\mu_{3}=\mu_{8}$ in (\ref{free-energy}), and work
close to the transition point $\delta \mu \gtrsim \delta \mu_{c}$
which is the point where $m_{M}^{2}$ continuously changes sign to a
negative value. At that small negative value of $m_{M}^{2}$ the
gluon condensate and the induced magnetic field should be very small
so simplifying the calculations. Note the difference between (\ref{free-energy}) and (\ref{Gibbs}). In (\ref{free-energy}) there is no applied field $\widetilde{H}$, but only and induced field $\widetilde{B}$.

Minimizing (\ref{free-energy}) with respect to $\overline{G}^{*}$
gives
\begin{equation}
\label{G-Eq} -\widetilde{\Pi}^{2}
\overline{G}-(2\widetilde{q}\widetilde{B}+|m_{M}^{2}|)\overline{G}+2g^{2}|\overline{G}|^{2}\overline{G}=0
\end{equation}
Eq. (\ref{G-Eq}) is a highly non-linear differential equation that
can be exactly solved only by numerical methods. Nevertheless, we
can take advantage of working near the transition point, where we
can manage to find an approximated solution that will lead to a
qualitative understanding of the new condensate phase. With this
aim, and guided by the previous experience with external fields \cite{Vortex, Olesen}, where the
solution is always such that the kinetic term
$|\widetilde{\Pi}_{\mu}K_{\nu}-\widetilde{\Pi}_{\nu}K_{\mu}|^{2}$ is
approximately zero near the transition point, we will consider that
when $\delta\mu \simeq \delta\mu_{c}$ our solution will satisfy the
same condition. Hence, we will look for a minimum solution
satisfying
\begin{equation}
\label{G-Eq-2} \widetilde{\Pi}^{2}
\overline{G}+\widetilde{q}\widetilde{B}\overline{G} \simeq 0.
\end{equation}
With the help of (\ref{G-Eq-2}) one can show that the minimum
equation for the induced field $\widetilde{B}$ takes the form
\begin{equation}
\label{B-Eq} 2\widetilde{q} |\overline{G}|^{2}-\widetilde{B}\simeq 0
\end{equation}
The relative sign between the two terms in Eq. (\ref{B-Eq}) implies
that for $|\overline{G}|\neq 0$ a magnetic field $\widetilde{B}$ is
induced. The origin of that possibility can be traced back to the
anomalous magnetic moment term in the action of the charged gluons.
This effect has the same physical root as the paramagnetism found in
Ref. \cite{Vortex}; where contrary to what occurs in conventional
superconductivity, the resultant in-medium field $\widetilde{B}$
becomes stronger than the applied field $\widetilde{H}$ that
triggers the instability.

Using the minimum equations (\ref{G-Eq}) and (\ref{B-Eq}) in
(\ref{free-energy}), we obtain the condensation free-energy density
\begin{equation}
\label{F-min} \mathcal{\overline{F}}_{g} \simeq
-2(g^2-\widetilde{q}^2)|\overline{G}|^4
\end{equation}
The hierarchy between the strong ($g$) and the electromagnetic
($\widetilde{q}$) couplings implies that
$\mathcal{\overline{F}}_{c}< 0$. Therefore, although the induction
of a magnetic field always costs energy (as can be seen from the
positive first term in (\ref{free-energy})), the field interaction
with the gluon anomalous magnetic moment, produces a sufficiently
large negative contribution to compensate for the increase.
Consequently, as seen from (\ref{F-min}), the net effect of the
proposed condensates is to decrease the system free-energy density.

It follows from Eqs.(\ref{G-Eq})-(\ref{B-Eq}) that near the phase
transition point the inhomogeneity of the condensate solution should
be a small but nonzero correction to a leading constant term
\begin{equation}
\label{Constraint-3} |\overline{G}|^{2}\simeq
\Lambda_{g/\widetilde{q}} |m_{M}^{2}|/2\widetilde{q}^{2} +\mathcal
{O}(m_{M}^4)f(x,y),
\end{equation}
\begin{equation}
\label{Constraint-2} \widetilde{q}\widetilde{B}\simeq
\Lambda_{g/\widetilde{q}} |m_{M}^{2}|+\mathcal {O}(m_{M}^4)g(x,y).
\end{equation}
with $\Lambda_{g/\widetilde{q}}\equiv(g^{2}/\widetilde{q}^{2}-1)^{-1}$. The explicit form of the inhomogeneity in the region $r\ll \xi$, where $\xi^{2}\equiv
1/\Lambda_{g/\widetilde{q}}|m_{M}^{2}|$ can be found to be (see for details Ref. \cite{Gluon-C})
\begin{equation}
\label{Amplitude}
|\overline{G}|^2\simeq\frac{1}{2\widetilde{q}^2\xi^2}-\frac{r^2}{4\widetilde{q}^2\xi^4}
\end{equation}
The improved solution for $\widetilde{B}$ is found substituting
(\ref{Amplitude}) back into (\ref{B-Eq}). The induced field
$\widetilde{B}$ is homogeneous in the $z$-direction and
inhomogeneous in the $(x,y)$-plane.

One may wonder whether this inhomogeneous gluon condensate forms a
vortex state. To answer this question we can compare our results
with the case with external magnetic field \cite{Vortex}. For this
we should notice that the mathematical problem we have just solved
is formally similar to that where the instability is induced by a
weak external field.  This would be the situation when the $2SC$
system approaches the transition point from the stable side (real
magnetic mass) and the external magnetic field is slightly larger
than the positive mass square $\widetilde{H}\simeq
\widetilde{H}_{c}= m_{M}^{2} \ll 1$. We know that at large
$m_{M}^{2}$ the condensate solution is a crystalline array of
vortex-cells with cell's size $\sim \xi\ll1$. At smaller $m_{M}^{2}$
the lattice structure should remain, but with a larger separation
between cells, since in this case $\xi\gg1$. However, the use of a
linear approximation to solve the equations in this case only allows
to explore the solution inside an individual cell ($r\ll\xi$). This
is the same limitation we have in the linear approach followed in
the present paper. Therefore, we expect that when the nonlinear
equations will be solved, the vortex arrangement will be explicitly
manifested.

\section{Astrophysical connotations of magnetized CS}

In addition to the fact that the low-energy physics of a magnetize quark star will have peculiar characteristics that should be reflected in its transport properties and specifically in its cooling process, as pointed out above, here I want to analyze other consequences for compact stars of the results presented in this talk.

There are some criticisms related to the origin of the large magnetic fields that are abundant in astrophysical compact objects as magnetars. If it is accepted the standard explanation of the origin of the magnetar's large magnetic
field through a magnetohydrodynamic dynamo mechanism that amplifies a seed magnetic
field due to a rapidly rotating protoneutron star, then it will imply that the rotation should have a spin period $<3 ms$.
Nevertheless, this mechanism cannot explain all the features of the supernova remnants surrounding these objects
\cite{magnetar-criticism}.

As we have discussed in this talk, in color superconductors magnetic fields tend to be reinforced and even to be generated. It is natural to believe that if a color superconducting state is realized in the core of neutron stars, it should have some implications for the magnetic properties of such compact objects. Taking into account that at the moderate high densities that can exist in the cores of neutron stars the charged gluons' Meissner masses decrease from values of order $m_{g}$ to values which are close to zero, and that any magnetic field with values $\widetilde{H}>m_M^2$ can produce the spontaneous generation of vortices of charged gluons that enhance the existing magnetic field, it becomes natural to expect that CS can have something to do with the generation of the large magnetic fields observed in some stellar objects as magnetars.

The gluon vortices found in Ref. \cite{Vortex} and discussed here, could produce a magnetic field of the
order of the Meissner mass scale ($m_{g}$), which implies a magnitude in the range $\sim 10^{16}-10^{17} G$. Hence, the possibility of generating a
magnetic field of such a large magnitude in the core of a compact
star without relying on a magnetohydrodynamics effect, can be an
interesting alternative to address the main criticism
\cite{magnetar-criticism} to the observational conundrum of the
standard magnetar's paradigm \cite{Magnetars}. On the other hand, to
have a mechanism that associates the existence of high magnetic
fields with CS at moderate densities can serve to single out the
magnetars as the most probable astronomical objects for the
realization of this high-dense state of matter.

Finally, I want to stress that as found in \cite{SpinoneCFL}, in the MCFL phase the magnetization is significantly enhanced. This feature can also serve to determine whether the core of magnetars is made of color superconducting matter or hadronic matter, since, as known, hadronic matter has negligible magnetization even at strong fields \cite{hadronic}.

\end{document}